\documentclass{procs}

\usepackage{amsmath}
\usepackage{enumitem}

\def\be{\begin{equation}}
\def\ee{\end{equation}}
\def\bea{\begin{eqnarray}}
\def\eea{\end{eqnarray}}

\newcommand{\Photo}{}

\newcommand{\gev}{\,\, \mathrm{GeV}}
\newcommand\MHp{M_{H^\pm}}

\begin{document}
\begin{flushright}
\mbox{}
IFT--UAM/CSIC--20-021~\footnote{Talk presented at the International Workshop on Future Linear Colliders (LCWS2019), Sendai, Japan, 28 October-1 November, 2019. C19-10-28.}  \\
DESY 20--019
\end{flushright}
\vspace*{1cm}

\title{The ``96 GeV excess'' at the ILC}

\author{T. Biek\"otter}
\addresss{DESY, Notkestrasse 85,
D-22607 Hamburg, Germany}

\author{M. Chakraborti}
\addresss{IFT (UAM/CSIC), Universidad Aut\'onoma de Madrid, 
Cantoblanco, E-28048, Spain}

\author{S. Heinemeyer}
\address{IFT (UAM/CSIC), Universidad Aut\'onoma de Madrid
Cantoblanco, E-28048, Spain\\
Campus of International Excellence UAM+CSIC, Cantoblanco, E-28049,
Madrid, Spain\\
Instituto de F\'isica de Cantabria (CSIC-UC), E-39005 Santander, Spain}

\maketitle\abstracts{
The CMS collaboration reported an intriguing $\sim 3 \, \sigma$ (local) excess
at $96\;$GeV in the light Higgs-boson search in the diphoton decay mode.
This mass coincides with a $\sim 2 \, \sigma$ (local) excess in the
$b\bar b$ final state at LEP. We present the interpretation of this
possible signal as the lightest Higgs boson in the 2 Higgs Doublet Model
with an additional real Higgs singlet (N2HDM). It is shown that the type II
and type IV (flipped) of the N2HDM can perfectly accommodate both
excesses simultaneously, while being in agreement with all
experimental and theoretical constraints. The excesses are most
easily accommodated in the type II N2HDM, which resembles the
Yukawa structure of supersymmetric models. We discuss the
experimental prospects for constraining our explanation at future
$e^+e^-$ colliders, with concrete analyses based on the ILC prospects.
}

\section{Introduction}

The Higgs boson discovered in 2012 by ATLAS and
CMS~\cite{Aad:2012tfa,Chatrchyan:2012xdj} is so far consistent with
the existence of a Standard-Model~(SM) Higgs
boson~\cite{Khachatryan:2016vau} with a mass of $\sim
125\,$GeV. However, the experimental uncertainties on the Higgs-boson
couplings are (if measured already) at the precision of $\sim 20\%$,
so that there is room for Beyond Standard-Model (BSM)
interpretations. Many theoretically well motivated extensions of the
SM contain additional Higgs bosons. In particular, the presence of
Higgs bosons lighter than $125\,$GeV is still possible. 

Searches for light Higgs bosons have been performed at LEP, the
Tevatron and the LHC. Besides the SM-like Higgs boson at $125\,$GeV no
further detections of scalar particles have been reported. However,
two excesses have been seen at LEP and the LHC at roughly the same
mass, hinting to a common origin of both excesses via a new particle
state. LEP observed a $2.3\, \sigma$ local excess 
in the~$e^+e^-\to Z(H\to b\bar{b})$
searches\,\cite{Barate:2003sz}, consistent with a
scalar of mass ~$\sim 98\,$GeV, where the mass
resolution is rather imprecise due to the
hadronic final state. The signal strength
was extracted to be $\mu_{\rm LEP}= 0.117 \pm 0.057$.
The signal strength $\mu_{\rm LEP}$ is the
measured cross section normalized to the SM expectation
assuming a SM Higgs-boson mass at the same mass.

CMS searched for light Higgs bosons in the diphoton
final state. Run II\,\cite{Sirunyan:2018aui} results
show a local excess of $\sim 3\, \sigma$ at
$\sim 96\,$GeV, and a similar excess of $2\, \sigma$
at roughly the same mass~\cite{CMS:2015ocq} in Run~I.
Assuming dominant gluon fusion production the
excess corresponds to $\mu_{\rm CMS}=0.6 \pm 0.2$.
First Run\,II~results from~ATLAS
with~$80$\,fb$^{-1}$ in the diphoton final state turned
out to be weaker than the corresponding CMS results, see, e.g., Fig.~1
in~\cite{Heinemeyer:2018wzl}.
Possibilities are discussed in the literature of how to
simultaneously explain both excesses by a common origin.
In particular supersymmetric realizations can be found in~\cite{Biekotter:2017xmf,Domingo:2018uim,Hollik:2018yek,Choi:2019jts,Biekotter:2019gtq,Cao:2019ofo}.
For a review we refer to Refs.~\cite{Heinemeyer:2018jcd,Heinemeyer:2018wzl},
see also Ref.~\cite{Richard:2020jfd}.

\section{The N2HDM}
We discussed in~\cite{Biekotter:2019kde,Biekotter:2019mib}
how a $\sim 96\,$GeV Higgs boson of the Next to minimal 2 Higgs
Doublet Model (N2HDM)~\cite{Chen:2013jvg,Muhlleitner:2016mzt} can be the
origin of both excesses in the type II and type IV scenarios.
The N2HDM extends the CP-conserving 2 Higgs Doublet Model (2HDM) by
a real scalar singlet field. In analogy to the 2HDM, a $Z_2$ symmetry
is imposed to avoid flavor changing neutral currents at the tree level,
which is only softly broken in the Higgs potential. Furthermore, a
second $Z_2$ symmetry, under which the singlet field changes the sign,
constraints the scalar potential. This symmetry is broken spontaneously
during electroweak symmetry breaking (EWSB), as soon as the singlet
field obtains a vacuum expectation value (vev).

In total, the Higgs sector of the N2HDM consists of 3 CP-even Higgs
bosons $h_i$, 1 CP-odd Higgs boson $A$, and 2 charged Higgs bosons
$H^\pm$. In principle, each of the particles $h_i$ can account for the
SM Higgs boson at $125\,$GeV. In our analysis, $h_2$ will be
identified with the SM Higgs boson, while $h_1$ plays the role of the
potential state at $\sim 96\,$GeV. The third CP-even and the CP-odd
states $h_3$ and $A$ were assumed to be heavier than $400\,$GeV to
avoid LHC constraints. The charged Higgs-boson mass was set to be
larger than $650\,$GeV to satisfy constraints from flavor physics
observables.

In the physical basis the 12 independent parameters
of the model are the mixing angles in the CP-even sector
$\alpha_{1,2,3}$, the ratio of the vevs of the Higgs doublets
$\tan\beta=v_2/v_1$, the SM vev $v=\sqrt{v_1^2+v_2^2}$, the vev of the
singlet field $v_S$, the masses of the physical Higgs bosons
$m_{h_{1,2,3}}$, $m_A$ and $M_{H^\pm}$, and the soft $Z_2$ breaking
parameter $m_{12}^2$. 
Using the public code
\texttt{ScannerS}~\cite{Coimbra:2013qq,Muhlleitner:2016mzt}
we performed a scan over the following parameter ranges:
\begin{align}
95 \gev \leq m_{h_1} \leq 98 \gev \; ,
\quad m_{h_2} = 125.09 \gev \; ,
\quad 400 \gev \leq m_{h_3} \leq 1000 \gev \; , \notag \\
400 \gev \leq m_A \leq 1000 \gev \; ,
\quad 650 \gev \leq \MHp \leq 1000 \gev \; , \notag\\
0.5 \leq \tan\beta \leq 4 \; ,
\quad 0 \leq m_{12}^2 \leq 10^6 \gev^2 \; ,
\quad 100 \gev \leq v_S \leq 1500 \gev \; . \label{eq:ranges}
\end{align}

\noindent
The following experimental and theoretical constraints were
taken into account:
\begin{itemize}[noitemsep,topsep=0pt]
	\item[-] tree-level perturbativity, boundedness-from-below
	and global-minimum conditions
	\item[-] Cross-section limits from collider searches
	using \texttt{HiggsBounds v.5.3.2}~\cite{Bechtle:2008jh,Bechtle:2011sb,Bechtle:2013wla,Bechtle:2015pma}
	\item[-] Signal-strength measurements of the SM Higgs boson
	using \texttt{HiggsSignals v.2.2.3}~\cite{Bechtle:2013xfa,Stal:2013hwa,Bechtle:2014ewa}
	\item[-] Various flavor physics observables, in particular 
excluding $\MHp < 650\gev$ for all values of 
$\tan\beta$ in the type~II and~IV.
	\item[-] Electroweak precision observables in terms
	of the oblique parameters $S$, $T$ and $U$~\cite{Peskin:1990zt,Peskin:1991sw}
\end{itemize}
For more details we refer to Ref.~\cite{Biekotter:2019kde}.
The relevant input for \texttt{HiggsBounds} and \texttt{HiggsSignals},
(decay withs, cross sections), were obtained using
the public codes
\texttt{N2HDECAY}~\cite{Muhlleitner:2016mzt,Djouadi:1997yw}
and
\texttt{SusHi}~\cite{Harlander:2012pb,Harlander:2016hcx}.

\section{Impications for future \boldmath{$e^+e^-$ colliders}}

The results of our parameter scans in the type II and type IV N2HDM,
as given in~\cite{Biekotter:2019kde}, show that both types of the
N2HDM can accommodate the excesses simultaneously, while being in
agreement with all considered constraints described above. A
preference of larger values of $\mu_{\rm CMS}$ in the type II scenario
is visible, which is caused by the suppression of decays into
$\tau$-pairs (see~\cite{Biekotter:2019kde} for details). 

The particle $h_1$ is dominantly singlet-like, and acquires its
coupling to the SM particles via the mixing with the SM Higgs boson
$h_2$. Thus, the presented scenario will be experimentally accessible
in two complementary ways. Firstly, the new particle $h_1$ can be
produced directly in collider experiments. Secondly, deviations of the
couplings of $125\,$GeV Higgs boson $h_2$ from the SM predictions are
present. We propose experimental analyses to constrain (or confirm)
our explanation of the excesses, both making use of the two effects
mentioned above. 

\subsection{Precision Higgs measurements: HL-LHC vs.\ ILC}

Due to the presence of the additional light Higgs boson
which is substantially mixed with the SM Higgs boson, 
the scenario deviates from the well-known alignment
limit of the 2HDM.
Currently, uncertainties on the measurement of the coupling
strengths of the SM-like Higgs boson at the LHC are still large,
i.e., at the $1 \, \sigma$-level they are of the same order as the
modifications of the couplings present
in our analysis in the
N2HDM~\cite{Khachatryan:2016vau,ATLAS-CONF-2018-031,Sirunyan:2018koj}.
In the future tighter constraints are expected from the LHC after the
high-luminosity upgrade (HL-LHC), when the planned amount
of $3000$\,fb$^{-1}$ integrated luminosity will have been
collected~\cite{Dawson:2013bba}. 
Finally, a future linear $e^+ e^-$ collider like the ILC could
improve the precision measurements of the Higgs boson couplings
even further~\cite{Dawson:2013bba,Drechsel:2018mgd}.\footnote{Similar
results can be obtained for CLIC, FCC-ee and CEPC. We will focus on the ILC 
prospects here using the results of Ref.~\cite{Drechsel:2018mgd}.} 
We compare our scan points to the expected precisions of
the LHC and the ILC as they are reported
in Refs.~\cite{Bambade:2019fyw,Cepeda:2019klc},
neglecting possible correlations of the coupling modifiers.

\begin{figure}
  \centering
  \includegraphics[width=0.60\textwidth]{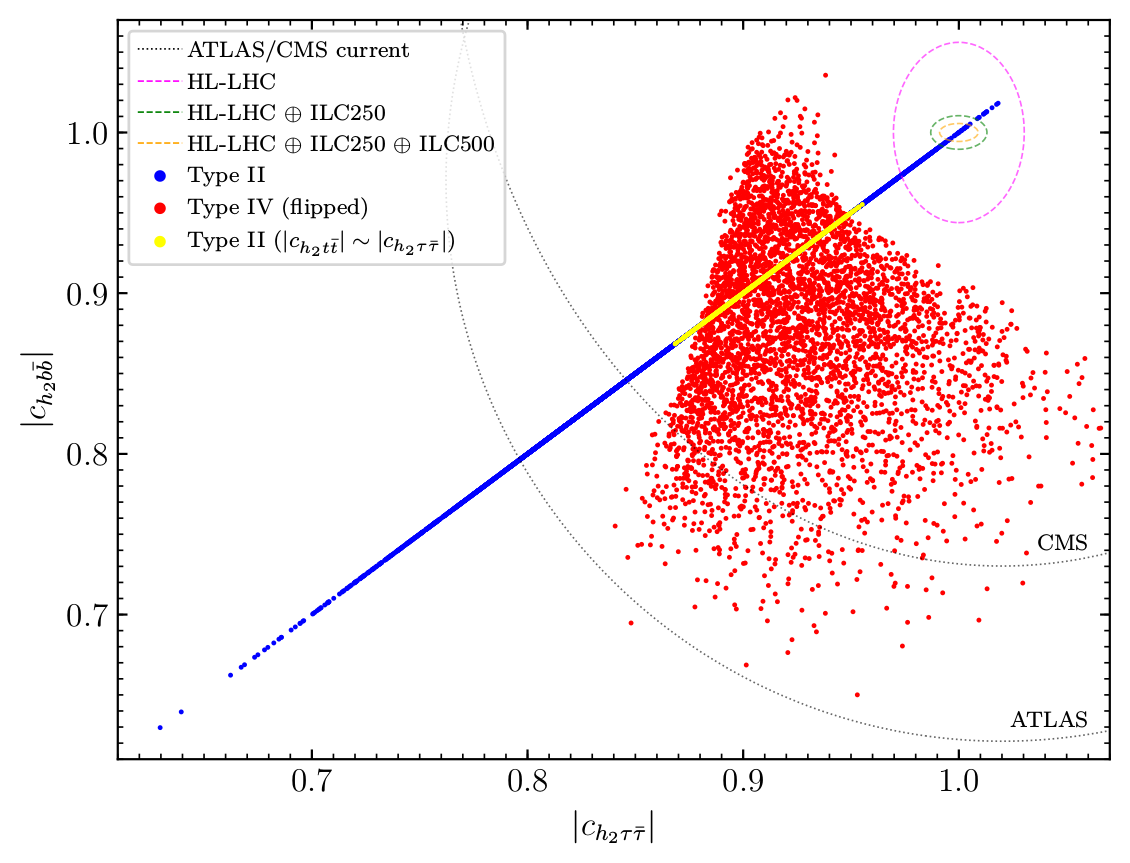}\\
  \includegraphics[width=0.60\textwidth]{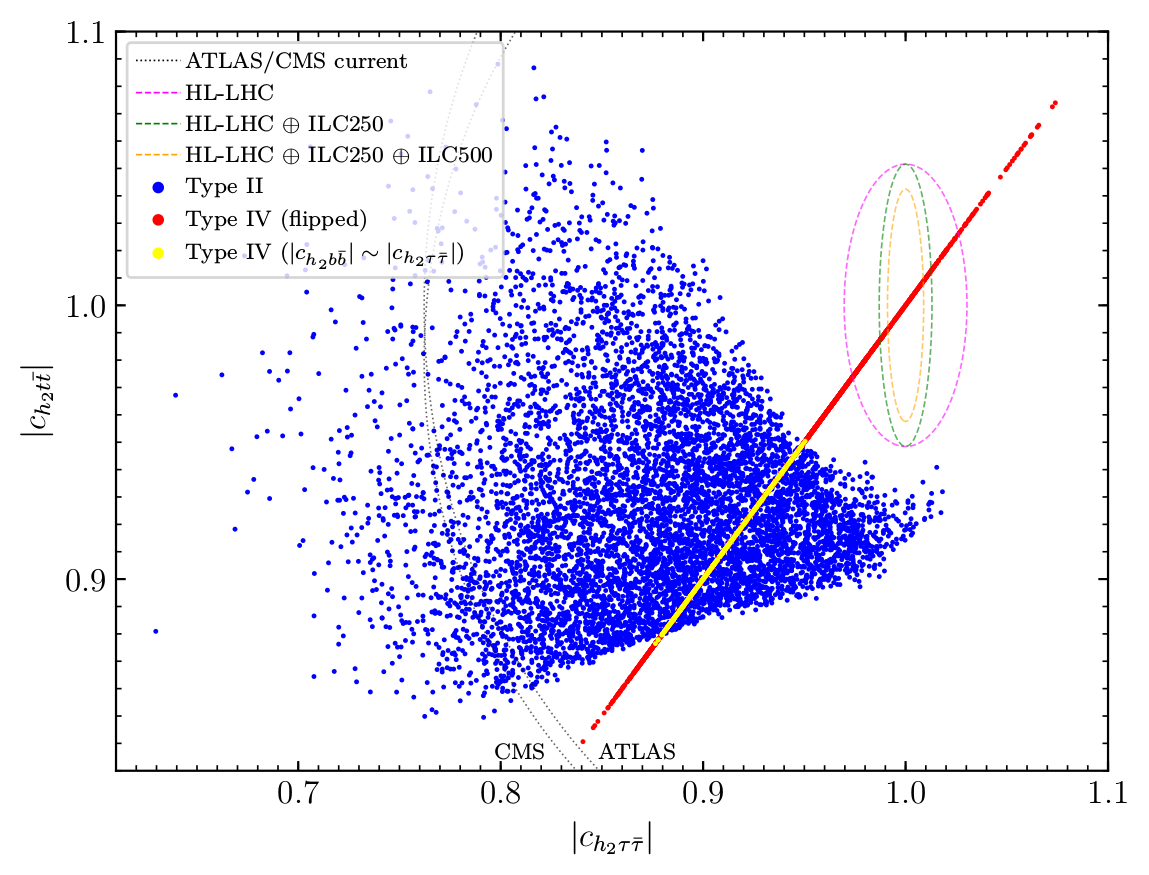}\\
  \includegraphics[width=0.60\textwidth]{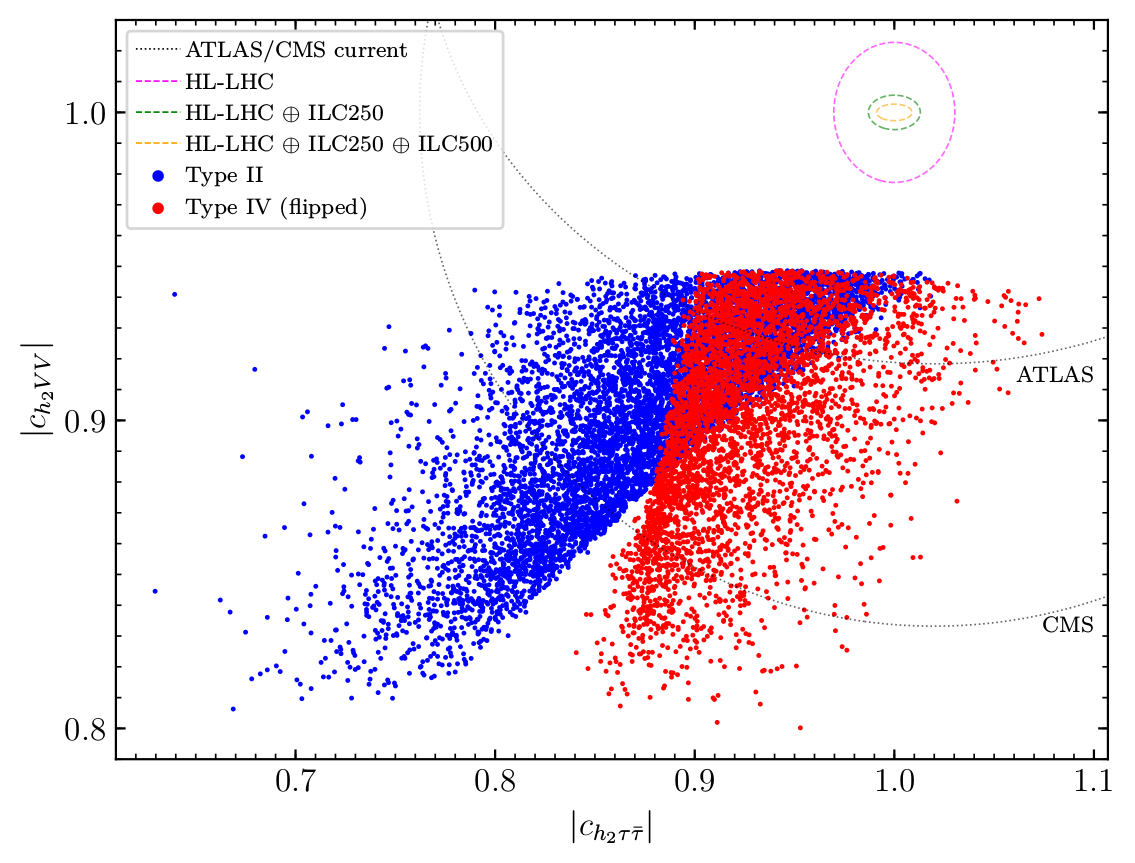}
  \caption{Prospects for the Higgs coupling measurements at the HL-LHC
  and the ILC (see text). The upper, middle and lower plot show the
  planes of $|c_{h_2\tau\tau}|-|c_{h_2t\bar t}|$,
  $|c_{h_2\tau\tau}|-|c_{h_2b\bar b}|$,$|c_{h_2\tau\tau}|-|c_{h_2VV}|$.}
  \label{fig:cplprosp}
\end{figure}

In Fig.~\ref{fig:cplprosp} we plot the coupling modifier
of the SM-like Higgs boson $h_2$ to $\tau$-leptons, $c_{h_2\tau\tau}$ 
on the horizontal axis against the coupling coefficient to
$b$-quarks, $c_{h_2b\bar b}$ (top), to $t$-quarks, $c_{h_2t\bar t}$
(middle) and to the massive SM gauge bosons, $c_{h_2VV}$ (bottom),
for both types.
These points passed all the experimental and theoretical constraints, including
the verification of SM-like Higgs boson properties in agreement
with LHC results using \texttt{HiggsSignals}.
In the top plot the blue points lie on a diagonal line, because in
type~II the coupling to leptons and to down-type quarks scale identically,
while in the bottom plot the red points representing
the type~IV scenario lie on the diagonal,
because there the lepton-coupling scales in the same
way as the coupling to up-type quarks.
The current measurements
on the coupling modifiers by ATLAS~\cite{ATLAS-CONF-2018-031}
and CMS~\cite{Sirunyan:2018koj} are shown
as black ellipses, although
the corresponding uncertainties are still very large.

We include several future precisions
for the coupling measurements. It should be noted that they are centered
around the SM predictions to show the potential to discriminate
the SM from the N2HDM. The magenta ellipse in each plot shows the
expected precision of the measurement of the coupling coefficients at
the $1 \, \sigma$-level at the HL-LHC from Ref.~\cite{Cepeda:2019klc}.
The current uncertainties and the HL-LHC analysis are based on the
coupling modifier, or $\kappa$-framework. These modifiers are then
constrained using a global fit to projected HL-LHC data assuming no
deviation from the SM prediction will be found. 
We use the uncertainties given under the assumptions that
no decay of the SM-like Higgs boson to BSM particles is present,
and that current systematic uncertainties will be reduced in addition
to the reduction of statistical uncertainties due to the increased statistics.

The green and the orange ellipses show the corresponding expected
uncertainties when the HL-LHC results are combined with projected
data from the ILC after the $250\gev$ phase and
the $500\gev$ phase, respectively, taken from Ref.~\cite{Bambade:2019fyw}.
Their analysis is based on a pure effective field theory calculation,
supplemented by further assumptions to facilitate the combination with
the HL-LHC projections in the $\kappa$-framework. In particular, in
the effective field theory approach the vector boson couplings can
be modified beyond a simple rescaling. This possibility was excluded
by recasting the fit setting two parameters related to the couplings
to the $Z$-boson and the $W$-boson to zero
(for details we refer to Ref.~\cite{Bambade:2019fyw}).

While current constraints on the SM-like Higgs-boson
properties allow for large deviations of the couplings of up
to $40\%$, the allowed parameter space of our scans will be significantly
reduced by the expected constraints from the HL-LHC
and the ILC.\footnote{Here one has to
keep in mind the theory input required in the (HL-)LHC analysis.}
For instance, the uncertainty of the coupling to $b$-quarks
will shrink below $4\%$ at the HL-LHC and below $1\%$ at the ILC.
For the coupling to $\tau$-leptons the uncertainty is expected to be
at $2\%$ at the HL-LHC. Again, the ILC could reduce this uncertainty further
to below $1\%$. For the coupling to $t$-quarks, on the other
hand, the ILC cannot substantially improve the expected uncertainty of
the HL-LHC (but permit a model-independent analysis).
Still, the HL-LHC and the ILC are
expected to reduce the uncertainty by roughly a factor of three.
This demonstrates that our explanation of the LEP
and the CMS excesses within the N2HDM is testable indirectly using future
precision measurements of the SM-like Higgs-boson couplings.

Comparing the top and middle plots in Fig.~\ref{fig:cplprosp} we find that,
independent of the type of the N2HDM, there is not a single benchmark
point that coincides with the SM prediction regarding the three
coupling coefficients shown. This implies that, once these
couplings are measured precisely by the HL-LHC and the ILC,
a deviation of the SM prediction has to be measured in at least one
of the couplings, if our explanation of the excesses is correct.
Conversely, if no deviation from the SM prediction regarding these
couplings will be measured, our explanation would be ruled out entirely.
Furthermore, in the case that a deviation from the SM prediction
will be found, 
the predicted scaling behavior of the coupling coefficients in the
type~II scenario (upper plot) and the type~IV scenario (middle plot),
might lead to distinct possibilities for the two models to accommodate these
possible deviations. In this case, precision measurements of the
SM-like Higgs boson couplings could be used differentiate between
the type~II and type~IV solution and thus to exclude one of the
two scenarios.
This is true for all points except the ones highlighted
in yellow in Fig.~\ref{fig:cplprosp}. The yellow points are a subset of points
of our scans that, if such deviations of the SM-like Higgs boson
couplings will be measured, could correspond to a benchmark point
both in the type~II and type~IV.

Finally, in the lower plot of Fig.~\ref{fig:cplprosp}, where the 
absolute value of the coupling modifier of the SM-like Higgs boson
w.r.t.\ the vector boson couplings $|c_{h_2 V V}|$ is shown on the
vertical axis, the parameter points of both 
types show deviations larger than the projected experimental uncertainty
at HL-LHC and ILC.
The deviations in $|c_{h_2VV}|$ are even stronger than for the
couplings to fermions. A $2\,\sigma$ deviation from the SM prediction
is expected 
with HL-LHC accuracy. At the ILC a deviation of more than $5\,\sigma$
would be visible. As mentioned already, a suppression of the
coupling to vector bosons is explicitly
expected by demanding $\Sigma_{h_2} \geq 10\%$. However,
since points with lower singlet component cannot accommodate
both excesses, this does not contradict the conclusion that
the explanation of both excesses can be
probed with high significance with future Higgs-boson coupling
measurements.


\subsection{Production of the 96 GeV Higgs at the ILC}

\begin{figure}
  \centering
  \includegraphics[width=0.80\textwidth]{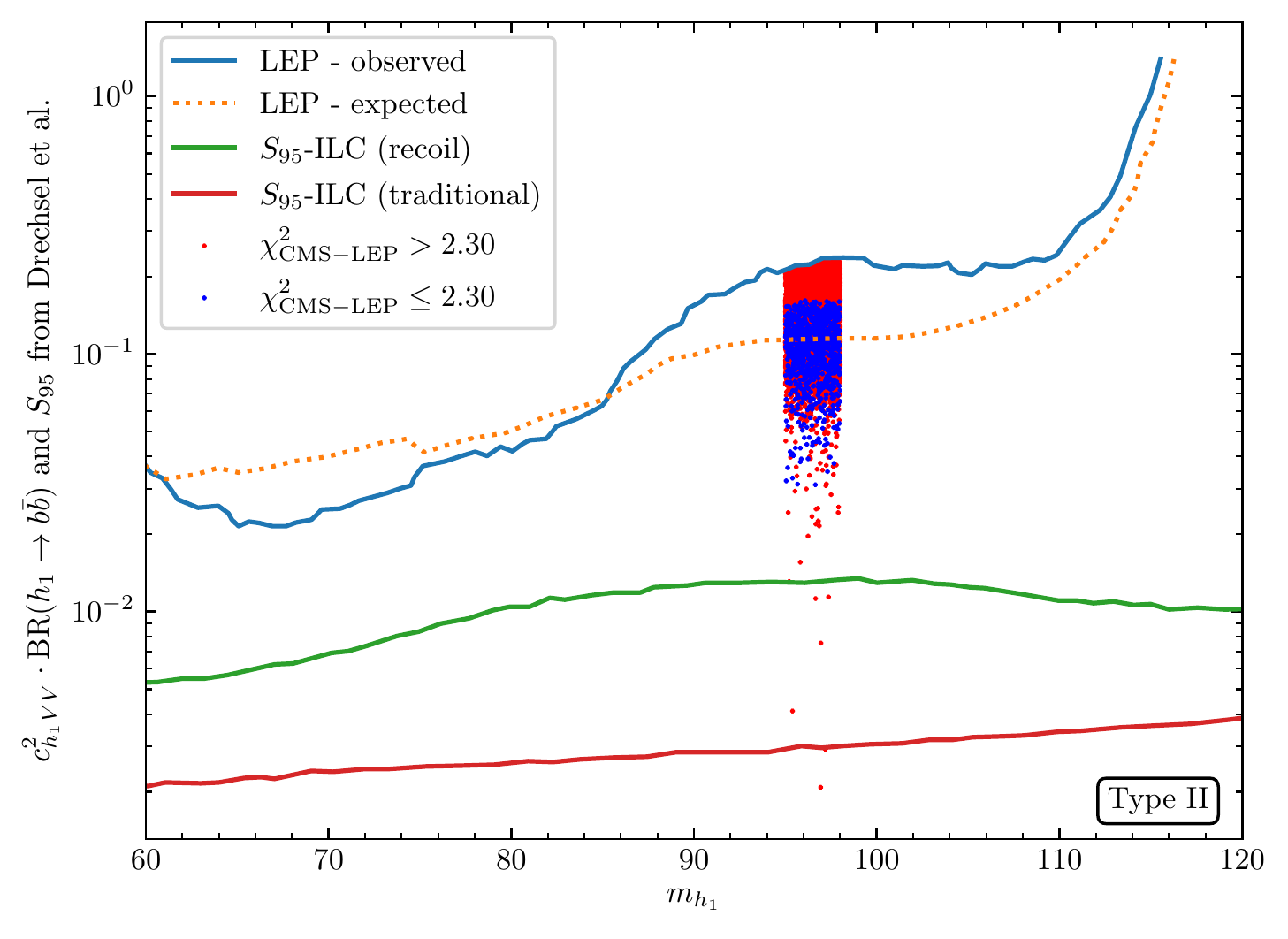}
  \caption{The $95\%$ CL expected (orange dashed) and
  observed (blue) upper bounds on the Higgsstrahlung production
  process with associated decay of the scalar to a pair of
  bottom quarks at LEP~\protect\cite{Barate:2003sz}.
  The expected ILC sensitivites are given by the green and red line
  (see text)~\protect\cite{Drechsel:2018mgd}. N2HDM type~II model
  points within (outside) the $1\,\sigma$~CL of the 96~GeV excess are
  shown as blue (red) dots. N2HDM type~IV results are similar.
  }
  \label{fig:2ilc}
\end{figure}

Regarding future collider experiments beyond the LHC, a
lepton collider is expected to be able to produce and
analyse the additional light Higgs boson $h_1$.
As an example, we compare
the current LEP bounds and the prospects of the
International Linear Collider (ILC), based on~\cite{Drechsel:2018mgd},
to our scan points 
in the type II scenario in Fig.~\ref{fig:2ilc}\,(left).
We show the expected $95\%$ CL upper limits at
the ILC using the traditional (red) and the
recoil technique (green)~\cite{Drechsel:2018mgd}. We indicate the points
  which lie within (blue) and
  outside (red) the $1 \, \sigma$ ellipse
  regarding $\mu_{\rm LEP}$ and $\mu_{\rm CMS}$.
Remarkably, all the points we found
that fit the LEP and the CMS excesses at the $1\,\sigma$
level would be excluded by the ILC, if no deviations from
the SM background would be observed. On the other hand,
if the $96\,$GeV Higgs boson is realized in nature, the
ILC would be able to produce it in large numbers.

\section*{Acknowledgements}

The work  was supported in part by the MEINCOP (Spain) under 
contract FPA2016-78022-P and in part by the AEI
through the grant IFT Centro de Excelencia Severo Ochoa SEV-2016-0597. 
The work of S.H.\ was 
supported in part by the Spanish Agencia Estatal de
Investigaci\'on (AEI), in part by
the EU Fondo Europeo de Desarrollo Regional (FEDER) through the project
FPA2016-78645-P, in part by the ``Spanish Red Consolider MultiDark''
FPA2017-90566-REDC.
T.B.\ is supported by the Deutsche Forschungsgemeinschaft under
Germany's Excellence Strategy EXC2121 ``Quantum Universe'' - 390833306.

\section*{References}
\bibliography{lcws-n2hdm}

\end{document}